\documentclass[final,5p,times,twocolumn]{elsarticle}

\usepackage{graphicx}

\usepackage{amssymb,amsmath,amsthm}

\usepackage{url}

\journal{Reliability Engineering and System Safety}

\begin{document}

\begin{frontmatter}



\title{Resource Redistribution under Lack of Information: \\ Short-Term Recovery after Large Scale Disasters}


\author{Vasily Lubashevskiy, Taro Kanno, and Kazuo Furuta}

\address{Department of Systems Innovations, School of Engineering, University of Tokyo, 
7-3-1 Hongo Bunkyo-ku, 113-8656 Tokyo, Japan}

\begin{abstract}
The paper is devoted to the problem of disaster mitigation. It develops an emergent mechanism of resource redistribution aimed at recovering of a socio-technological system affected by a large scale disaster. The basic requirements to the short-term recovery are taken into account in constructing this mechanism. The system at hand consists of many individual units (cities) and the mechanism is based on their cooperative interaction, which makes the resource redistribution efficient regardless of the particular position of affected region. Previously we studied the dynamics of supply process governed by this mechanism when all the information about the system is available and actual from the beginning of the process 
(V. Lubashevskiy, T. Kanno, K. Furuta, arXiv:1310.0648). 
In the present paper we analyze the effects of lack of information on the resource delivery rate. Two scenarios of the lack of information are allowed for. The first one is the delay of the information about the city states. The second one is its incompleteness. As a result of simulation, it is demonstrated that the duration of the resource redistribution governed by the developed mechanism is weakly affected by the lack of information. 

\end{abstract}


\end{frontmatter}


\section{Introduction}
\label{}

In recent years the problems of disaster mitigation and resilience have attracted much attention. As far as mitigation of large scale disasters is concerned, two phases, short- and long-term recoveries, can be distinguished. The use of these terms has a long history \cite{national1979comprehensive}, nonetheless, the appropriate classification of recovery phases is required especially for efficient emergency management of large scale disasters \cite{DHS2008,MalcolmBaird2010a,DHS2013}.   

Following the cited materials we consider the short-term recovery \textit{mainly} aiming at restoring the vital life-support system to the minimal operating standards. Generally this system comprises many individual components and the corresponding services, in particular, sheltering, feeding operations, emergency first aid, bulk distribution of emergency items, and collecting and providing information on victims to family members. One of the basic requirements imposed on the short-term recovery is the beginning of its implementation within the minimal time. For example, the aforementioned services have to start their operations within 8 hours according to the Disaster Recovery Plan of 
State of Illinois \cite{DRP_Illinois_July2012_STRecovery}.

To mitigate aftermath of a large scale disaster cooperation of many cities is required because the amount of resources initially accumulated in an affected area can be insufficient to recover all the individual components of the vital life-support system. Thereby the implementation of the short-term recovery is directly related to an efficient resource redistribution. This redistribution cannot be predetermined because of the unpredictability of disaster consequences. It is possible only to formulate rather general requirements for this process. First, the supply to an affected area must start practically immediately in order to recover the life-support system. Second, the resource supply should be decentralized, otherwise, its centralized management can be a `bottleneck' that delays the responsive and adaptive delivery of resources or aid \cite{managprob:panda2012}. 

In the previous work \cite{lubashevskiy2013algorithm} we have proposed an algorithm by which the required resource redistribution can be implemented in a quasi-optimal way. Initially it was applied to modeling the resource redistribution in an affected system for which the whole information about the city states is available from the beginning of the short-term recovery. In the reality one of the crucial points for the recovery management is  lack of information about the affected area just after the disaster. The purpose of the present paper is to analyze the effects of lack of information on the recover rate.

Two scenarios, namely, the delay of information and its incompleteness will be considered. The former scenario implies that to collect the whole information some time is required, so, the real situation may be much worse than the initial evaluation of the disaster consequences. The latter scenario represents a more problematic situation. For instance, in the case of earthquakes in mountainous area the communication with some cities can be broken  \cite{managprob:panda2012} and the information about their state is just not available at the beginning of the recovery process. 

One of the key points of the present work is that the developed algorithm generates a plan for the resource redistribution based on the current information about the city states. If during the implementation of this plan the information is updated, the plan will be reconstructed without halting the process remarkably and the resource redistribution will be continued taking into account the changed information. As will be demonstrated, this approach, in particular, gives rise to a faster implementation of the short-term recovery than if it would be postponed until the whole information is collected.

\section{Model}


\subsection{Model background}\label{modback}

The Great East Japan Earthquake occurred along the eastern coast of Japan on 11 March 2011 exemplifies large scale destructive disasters that necessitate the cooperation of many cities and even regions in mitigating the aftermath. The hypo-central region of this earthquake comprised several offshore prefectures (Iwate, Miyagi, Fukushima, and Ibaraki Prefectures) and have  ruptured the zone with a length of 500~km and a width of 200~km \cite{wiki:GEJ2011}. The terrible aftermath of the disaster initiated evacuation from some areas of these prefectures, thousands houses were destroyed, many victims required medical assistance. Obviously none of the affected cities was able to recover only by its local resources, practically all the non-affected cities in these prefectures were involved into this process. These Japanese prefectures can be one of the representative examples of the system, which overcame the disaster and recovered to its normal state. In numerical simulation to be described below some of the system parameters were evaluated comparing  with the data of the Fukushima prefecture. Namely, the total number of residents is evaluated as $4*10^6$, the area of the region treated as a certain administrative unit responsible for mitigating the aftermath is set about $4*10^5$~km$^2$, the mean distance between the neighboring cities in this region is 40--50~km, as a results, the number of cities that can be involved into recovering the affected region may be about 81. Such parameters correspond to the administrative unit comparable with two Fukushima prefectures.

\subsection{System under consideration}

The system is modeled as a collection of cities connected with one another by a transport network. Initially in each city $i$ there is some amount of the vital resources $Q_i$. Under the normal conditions this amount of resources is excessive and substantially exceeds the minimal critical level $Q_{ci}$ required for its residents to survive during a certain period of time, $Q_i > Q_{ci}$. Naturally the magnitude of the quantity $Q_{ci}$ depends on the number $N_i$ of residents in a given city $i$; the larger the number of residents, the higher the required level of resources $Q_{ci}$. One of the consequences of a large scale disaster is the increased demand for the vital resources in the affected cities. This is modeled as the essential increase in the corresponding magnitudes of $Q_{ci}$ and the opposite inequality $Q_i < Q_{ci}$ holds for the affected cities, which is the mathematical description of the disaster effect. Naturally the inequality
\begin{equation}\label{in:1}
\sum_i Q_i > \sum_i Q_{ci}
\end{equation}
must hold still after the disaster. Actually inequality (\ref{in:1}) is the mathematical implementation of the requirement that the given system is capable to survive as a whole during a certain time without external help.

\begin{figure}
\begin{center}
\includegraphics[width=0.8\columnwidth]{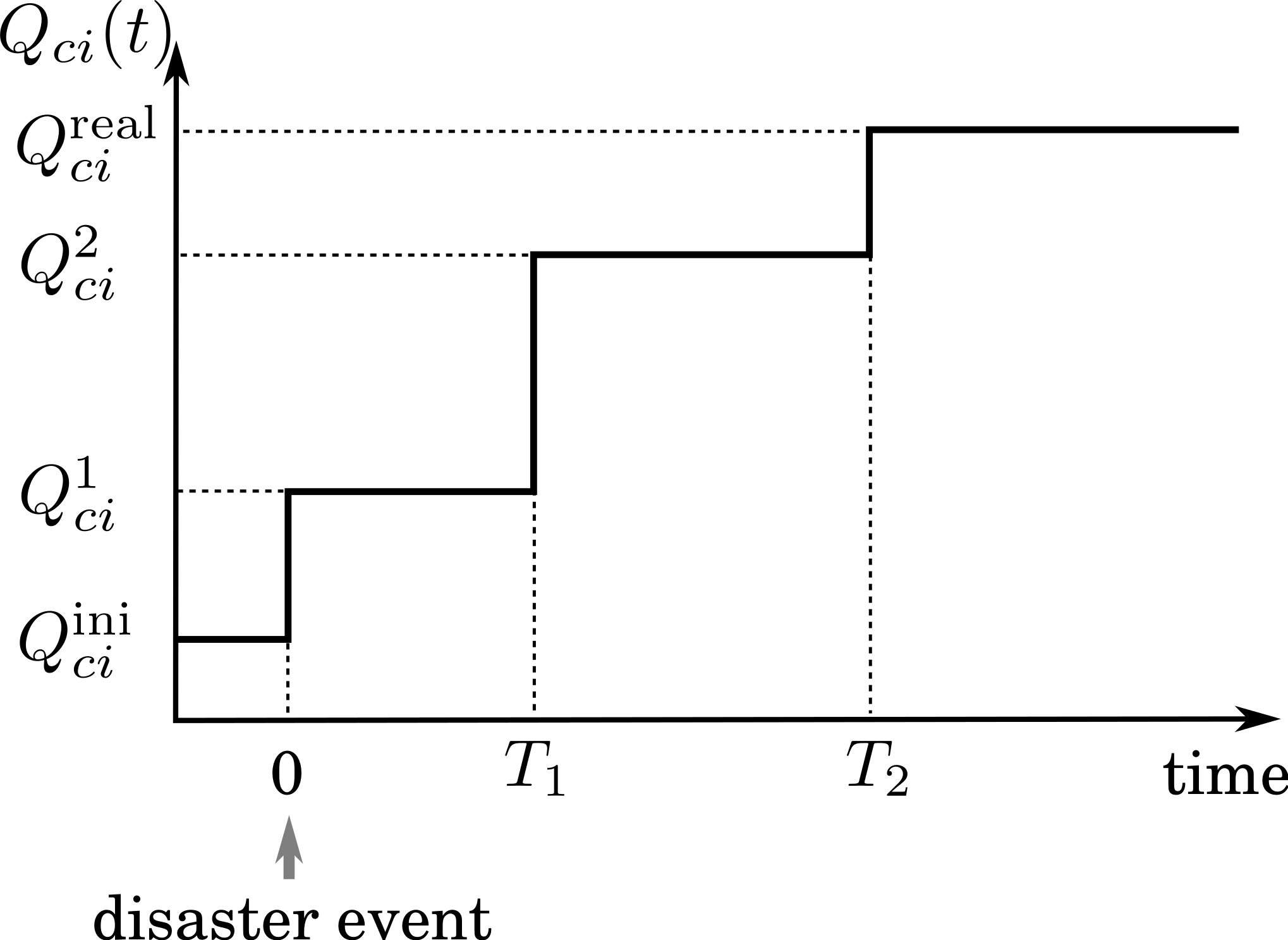}
\end{center}
\caption{Dynamics of the information update about the state of an affected city $i$ used in the simulation of the resource redistribution in the case of information delay.}
\label{F1}
\end{figure}

The effects of the delay and incompleteness of the information about the city states are analyzed individually. 

The effect of the information delay is described within the following model (Fig.~\ref{F1}). Just after the disaster the demand for the vital resources increases from the initial level $Q^\text{ini}_{ci}$ to the level $Q^\text{real}_{ci}$ in a given affected city $i$. However, the information about this city state is collected with a delay. The model at hand considers two steps of the information update. Just after the disaster the demand is evaluated to equal to $Q_{ci}^1$, then in the time $T_1$ the first update of information occurs and the level raises to $Q_{ci}^2$, and only at the time $T_2$ the real information is obtained.    

\begin{figure}
\begin{center}
\includegraphics[width=0.8\columnwidth]{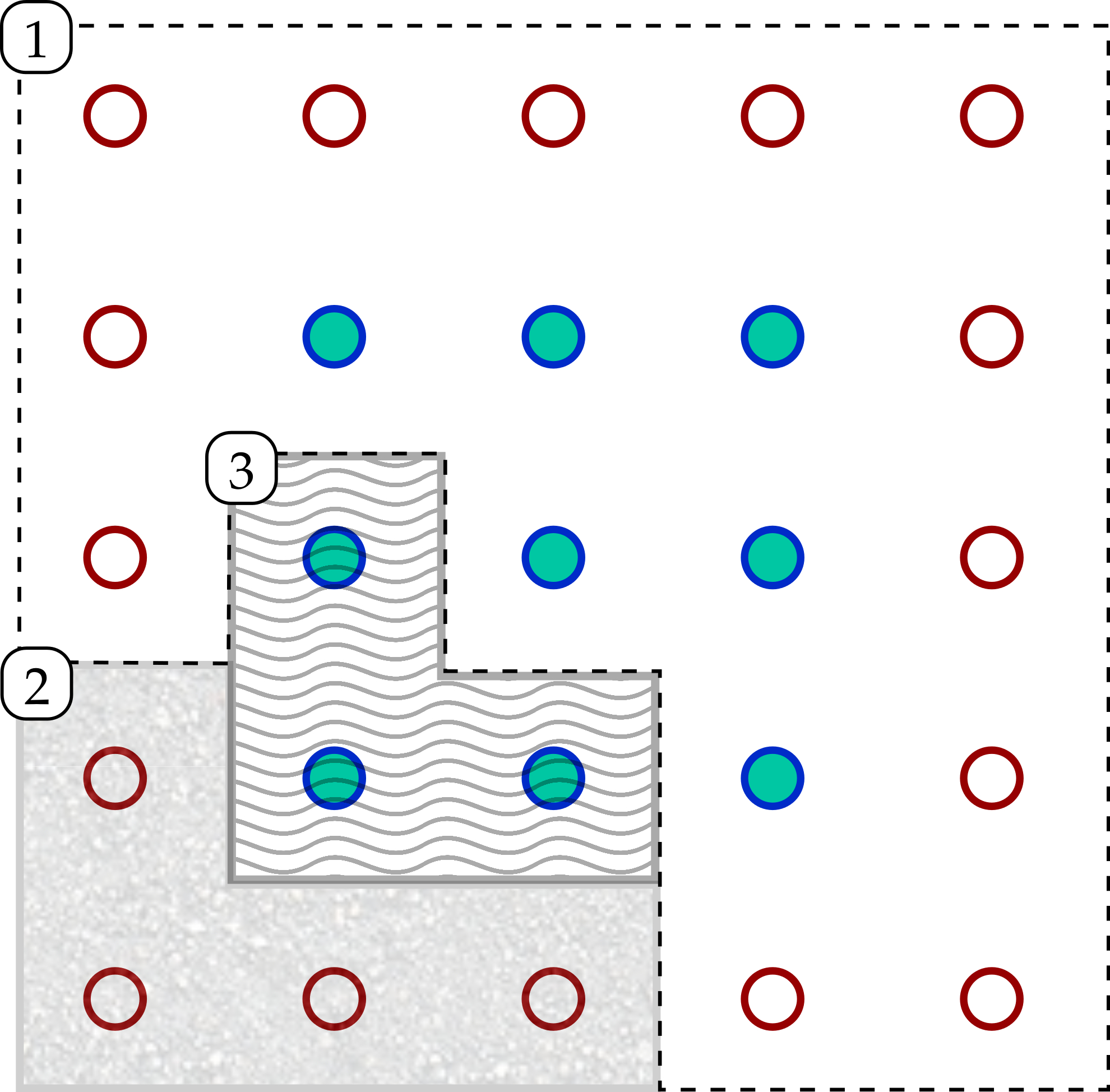}
\end{center}
\caption{Fragment of the system. Empty (red) circles represent the non-affected cities for which $Q_{ci} < Q_i$, filled (blue) circles depict the affected cities, $Q_{ci} > Q_i$. Region 1 is the collection of the cities for which the information about their state is available at the initial stage, the information about the state of the cities located inside regions 2 and 3 becomes available after the time $T_1$ and $T_2$, respectively.}
\label{F2}
\end{figure}

The effect of the information incompleteness is modeled as demonstrated in Fig.~\ref{F2}. The system is divided into three regions. The state of all the cities located in region 1 is assumed to be known right after the disaster. The information about regions 2 and 3 becomes available at the time moments $T_1$ and $T_2$, respectively. 

It should be noted that if the intervals $T_1$ and $T_2$ were known beforehand to be rather short and the duration of the process were not crucial, it would make sense to wait until all the information about the system state is collected. However, if the time of information updates is not determined and the recovery has to be initiated immediately, resource redistribution must be started as soon as possible regardless of the lack of information.

\section{Resource redistribution algorithm}

At the initial step all the cities that are accessible provide the information about their state, namely, the available amount of resources $Q_i$, the minimal critical amount $Q_{ci}$ required for their individual surviving, and the number $N_i$ of citizens. The characteristics of the transportation network are assumed to be also given, it is the matrix $\mathbf{D}=\|d_{ij}\|$ whose element, e.g., $d_{ij}$ specifies the minimal time distance between city $i$ and $j$. To describe the states of cities let us introduce the value
\begin{equation}\label{eq:theta}
	\theta_i = \frac{Q_i - Q_{ci}}{Q_{ci}}\,.
\end{equation}
If the information about a given city $i$ is not available, then the corresponding value is set equal to zero, $\theta_{i}=0$. When $\theta_i < 0$ its magnitude quantifies the lack of vital resources in relative units. The quantity $S_i = \theta_i N_i$, or more strictly its absolute value is actually the number of people being under the level of surviving. It will be used in specifying the priority of the cities in the resource redistribution queue.  
As previously \cite{lubashevskiy2013algorithm} to avoid the discussion about ethics or morality of the priority choice we appeal to the following historical example.  Baron Dominique Jean Larrey, surgeon-in-chief to Napoleon's Imperial Guard, articulated one of the first triage rule in 1792: ``Those who are dangerously wounded should receive the first attention, without regard to rank or distinction. They who are injured in a less degree may wait until their brethren in arms, who are badly mutilated, have been operated on and dressed, otherwise the latter would not survive many hours; rarely, until the succeeding day'' \cite{Napaleon}. The minimal value of $S$ corresponds to the maximal number of residents which are not supplied with the vital resources and it allows us to regard that city as most ``dangerously wounded''.

Because the main goal of resource redistribution just after the disaster is mitigation of consequences and minimization of the amount of victims, Table~\ref{T1} determines the priority of the resource redistribution.
\begin{table}[h]
\caption{The order of cities according to the resource redistribution priority. Here $M$ is the total number of cities in the given system.}\label{T1}
\begin{center}
\begin{tabular}{|c||c|c|c|c|c|}
\hline
$S_p$ & $S_1$ & $S_2$ & $\ldots$ & $S_{M-1}$ & $S_{M}$ \\
\hline
$p$ &   $1$ &  $2$ &  $\ldots$ &    $M-1$ &  $M$  \\
\hline 
\end{tabular}
\end{center}
\end{table}
\noindent
The order used in Table~\ref{T1} matches the inequality
\begin{equation}\label{eq:order}
S_1 \leq S_2 \leq \ldots\leq S_{M-1}\leq S_{M}
\end{equation}
and $i_1$, $i_2$, $\ldots$, are the corresponding indexes of these cities. 

In order to describe resource redistribution dynamics, let us introduce the following quantities. First, it is a certain quantum $h$ of resources that can be directed from a city to another one. The second quantity is the time $\Delta t$ required for this quantum to be assembled for transportation. The third one  $c_i$ is the capacity of a given city $i$ specifying the maximal number of quanta which can be assembled during the time $\Delta t$. Introduction of these quantities implies the realization of resource redistribution mainly via fast loading vehicles, for instance, tracks. In this case $h$ is the volume of resources transported by the typical vehicle individually, $\Delta t$ is the time necessary to load this vehicle, and $c_i$ is determined by the number of loading places and the capacity of loading facilities. 

The algorithm to be described below creates a complete plan of resource redistribution depending explicitly on the initial post-disaster system state. Namely, at the first step Table~\ref{T1} is formed using the initial data. The city $i_1$ is selected as the city with the wost situation. Then we choose a city $i_k$ such that
\begin{equation}\label{eq:dsearch}
d_{i_1 i_k} = \min_j d_{i_1 j}\quad \text{among} \quad Q_j - h \geq Q_{cj}\,.
\end{equation}
Then the prepared quantum is \textit{virtually} transported to city $i_1$ from city $i_k$ . It gives rise to the transformations
\begin{equation}\label{eq:transf}
\begin{split}
Q_{i_1} &\rightarrow Q_{i_1} + h\,,\\ 
Q_{i_k} &\rightarrow Q_{i_k} - h\,, \\ 
c_{i_k} & \rightarrow c_{i_k} - 1\,.
\end{split}
\end{equation}
The information about the given action is saved as a report of its virtual realization and comprises: ``city $i_k$ sent one quantum to city $i_1$ at time $t_{\text{dep},i_k}$, the quantum is received at time $t_\text{arr}$''. 
Initially for all the cities involved in the resource redistribution we set $t^\text{init}_{\text{dep},i_k}=0$. The further modification of these values will be explained below, see Eq.~\eqref{eq:decrD}.
In the developed algorithm the time moments $\{t_\text{arr}\}$ are specified via the expression
\begin{equation}\label{eq:tdeptarr}
t_\text{arr} =  d_{i_1 i_k}\,. 
\end{equation}
It should be noted that formula~\eqref{eq:tdeptarr} obviously holds at the initial steps when $t^\text{init}_{\text{dep},i_k}=0$ and the original element of the matrix $\mathbf{D}$ enters it. Its use in the general case will be justified by the renormalization of the matrix  $\mathbf{D}$ (see Eq.~\eqref{eq:decrD}) taking into account the delay in sending the resource quanta caused by the city limit capacity leading to nonlinear effects in the resource redistribution.

At the next step this procedure is reproduced again. Table~\ref{T1} is reconstructed, the logic of choosing the interacting cities is repeated with saving the relevant report. 

Since the maximal number of quanta that can be sent from a given city simultaneously is finite, there exist a situation where $c_j$ takes a zero value due to transformations~\eqref{eq:transf}. This effect is taken into account by renormalization of the matrix $\mathbf{D}$, which is a time distance matrix. Namely, when $c_j = 0$ we restore the initial value of $c_j$ and for all $i$
\begin{equation}\label{eq:decrD}
\begin{split}
d_{ij} &\rightarrow  d_{ij} + \Delta t\,,\\ 
t_{\text{dep},j} &\rightarrow t_{\text{dep},j}+ \Delta t\,.
\end{split}
\end{equation}
This procedure is terminated when at the next step 
\begin{equation}
\forall i: \quad S_i > 0\,.
\end{equation}
As a result, this algorithm generates the collection of reports which enables us to create a semi-optimal plan of resource redistribution for all the cities and the real process of resource redistribution is initiated.

According to this plan the cities start sending the real resources. If at a certain moment of time $T_1$ new information about the system state is received the procedure of plan construction is repeated. This reconstruction takes into account two factors. First, it is the new data about the city damage $\{Q_{ci}\}$. The second factor is the current pattern of resource allocation in the system, $\{Q_{i}(T_1)\}$. It is determined by the implementation of the previous plan of sending the resource quanta before the moment $T_1$. Within the frameworks of the developed algorithm this construction is possible because it is based on the collection of reports like ``city $i_k$ sent one quantum to city $i_1$ at time $t_\text{dep}$, the quantum is received at time $t_\text{arr}$'' and at any time moment it is possible to figure out how many resource quanta have been sent, got the destination, and are in the transportation process. Thereby, the new plan replaces the previous one from the time moment $T_1$ and continues governing the further resource redistribution. In the case of a new update event such reconstruction is repeated again.

This procedure of the plan reconstruction with the information update is a main modification of the algorithm developed in our previous work  \cite{lubashevskiy2013algorithm}. It enables us to govern the resource redistribution efficiently even in the case of the lack of information.

\section{Results of Numerical Simulation}


\begin{figure}
\begin{center}
\includegraphics[width=0.5\columnwidth]{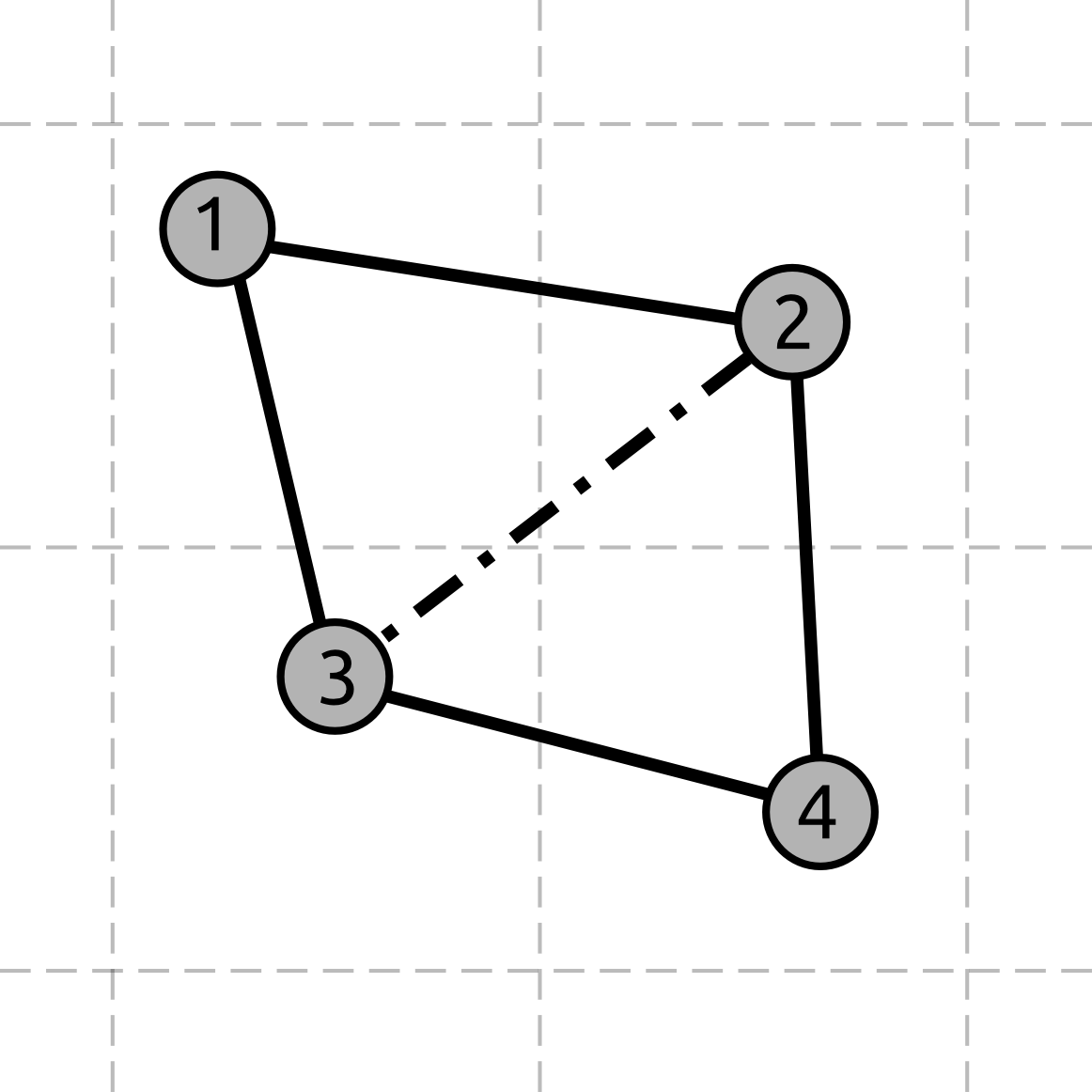}
\end{center}
\caption{Example of the city arrangement and the corresponding transportation network.}
\label{F:angle}
\end{figure}

\subsection{Details of modeling}\label{moddetail}

The purpose of the present section is to illustrate the  characteristic features of the analyzed resource redistribution process. Keeping in mind the administrative units noted in Section~\ref{modback}, the following system was studied numerically based on the developed model. It is assumed to comprise $9\times 9 = 81$ cities regarded as basic entities connected with one another by a transport network. The total population of these cities is set equal to $N\approx4\times 10^6$. The amount of resources were measured in the units of resource quantum $h$, so we set $h=1$. To be specific the volume of one quantum is assumed to supply 20 residents with the resources under the normal conditions. So the integral amount of resources initially allocated in the system is 
\begin{equation*}
\sum_i Q_i = \frac{N}{20}\,.
\end{equation*}
The mean time distance between the cities varied between 40 and 80 minutes and the time $\Delta t$ necessary to assemble one resource quantum was set equal to 1 minute. The number of quanta $h$, which could be formed in any city simultaneously was set equal to 1. This assumption was accepted to make the dynamics of supply as smooth as possible. It should be noted that in the previous paper \cite{lubashevskiy2013algorithm} we analyzed the resource redistribution supposing that 15 quanta can be assembled simultaneously within 15 minutes. Given the other conditions equal, it leads actually to the same dynamics, where, however, the effects of discretization are much more pronounced.     

The transportation network was constructed in the following way. The region occupied by the given system is considered to be of a rectangular form and is divided into $81$ (the number of cities) equal rectangles. Each rectangle contains one city placed randomly within it. At the first step the connections between the cities located in the neighboring rectangles are formed as illustrated in Fig.~\ref{F:angle}. For any arrangement of these four cities the ``vertical'' and ``horizontal'' connections are formed. A diagonal connection, for example, the connection 2-3 is formed if both of the opposite angles are less than $90^\text{o}$: $\angle{213}$ and $\angle{243}$ in Fig.~\ref{F:angle}. In this way we construct the matrix \textbf{D} of minimal time distances between the neighboring cities. The relationship between spatial and temporal scales was determined assuming the average speed of transporting vehicles to be equal to 60 km/h. At the next step using Warshall's algorithm (see, e.g., \cite{discmathstr_kbr6}) we completed \textbf{D} to the matrix of the minimal time distances between any pair of cities.

The number of the affected cities was chosen equal to 9 and they were located in the middle part the system in order to avoid the influence of possible edge effects. Before ascribing particular values of the time moments $T_1$ and $T_2$  we simulated the system recovery provided all the information is available. It gave us the time $T$ at which the last resource quantum was sent and enables us to evaluate the duration of the short-time recovery under ideal conditions. Then the time moments of the information update was set equal to $T_1 = 0.2\,T$ and $T_2 = 0.4\, T$.    

\subsection{Scenario of information delay}

Let us, first, present the results for the system with the information delay. It was assumed (Fig.~\ref{F1}) that at the beginning of the process (before $T_1$) the demand in three of nine affected cities is evaluated  as $Q_{ci}^1 = 0.5\,Q_{ci}^\text{real}$; during the time interval $T_1<t<T_2$ the demand is considered to be equal to $Q^2_{ci} = 0.8\,Q_{ci}^\text{real}$, and only after the second update at time $T_2$ the real value $Q_{ci}^\text{real}$ becomes known. For the affected cities the value of $Q_{ci}^\text{real}$ was set equal to $Q_{ci}^\text{real} = 3Q^\text{init}_i$ whereas for the unaffected cities this value was equal to $Q_{ci}=0.6 Q^\text{init}_i$, where $Q^\text{init}_i$ is the amount of resources under the normal conditions.  

\begin{figure}
\begin{center}
\includegraphics[width=1.0\columnwidth]{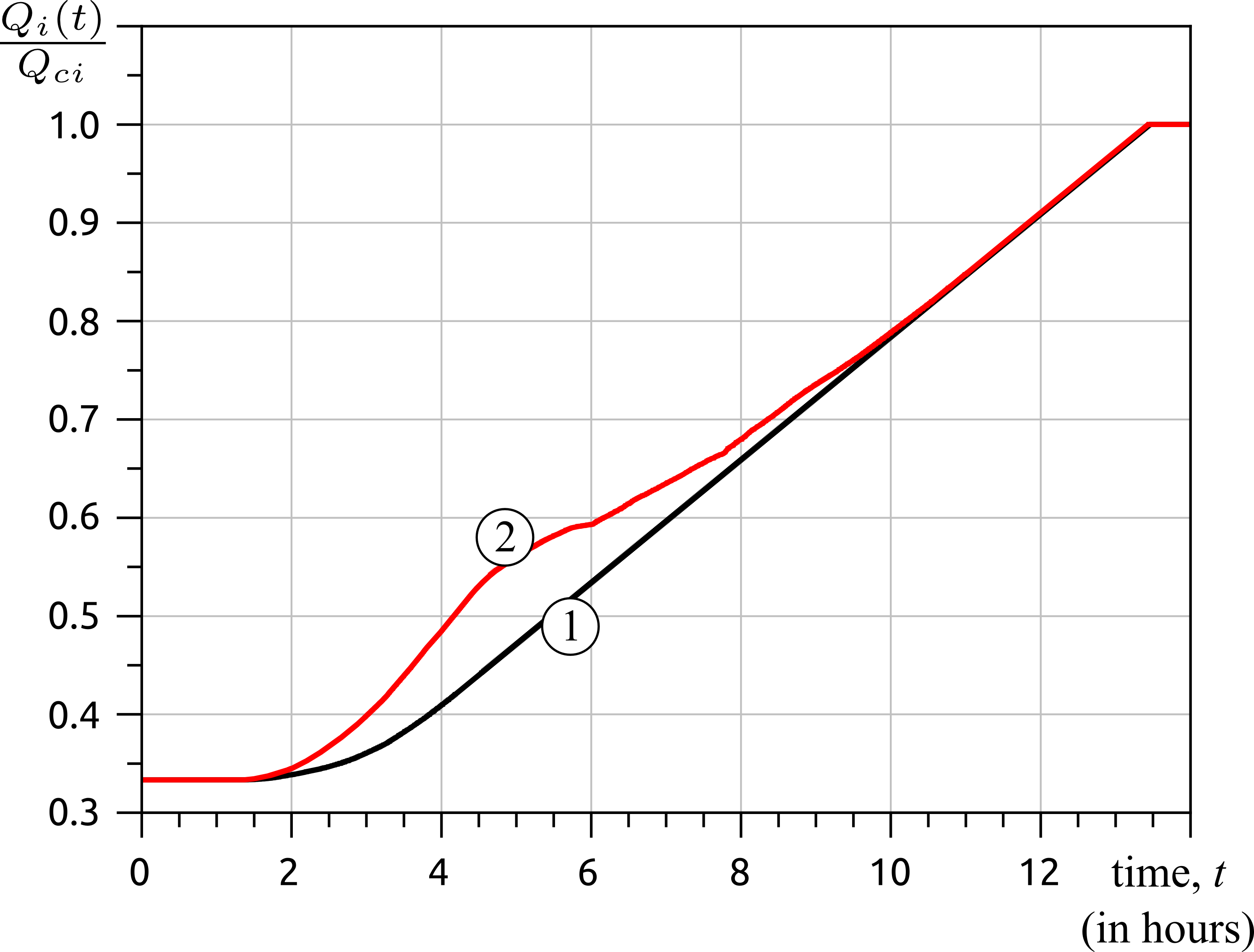}
\end{center}
\caption{Supply dynamics of an affected city $i$ that provided the sufficient information about its state from the beginning of the resource redistribution in two situations. Curve 1 matches to the ideal case, curve 2  corresponds to the scenario of information delay.}
\label{delay1}
\end{figure}

\begin{figure}
\begin{center}
\includegraphics[width=1.0\columnwidth]{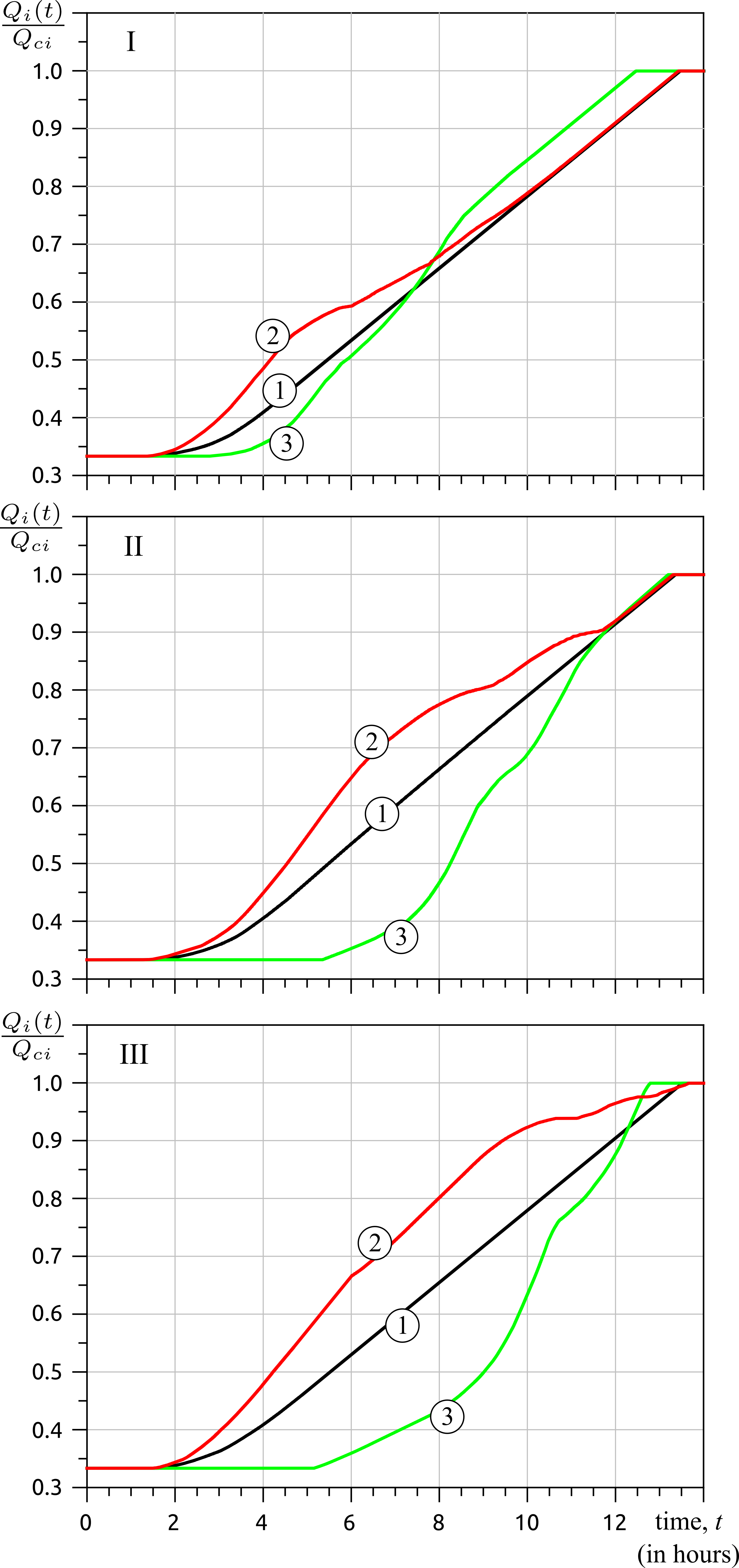}
\end{center}
\caption{Supply dynamics of affected cities in the scenario of information delay. Curves 1 corresponds to the ideal case, curves 2 to a city $i$ that provided the sufficient information about its state from the beginning of the resource redistribution, curves 3 to a city $j$ that provides the information about its state with delay. Plot I corresponds to $T_1 = 0.2\,T\approx 2.5~\text{h}$, $T_2 = 0.4\,T\approx 5~\text{h}$, Plot II corresponds to $T_1 = 0.4\,T\approx 5~\text{h}$, $T_2 = 0.6\,T\approx 7.5~\text{h}$, and Plot III corresponds to $T_1 = 0.6\,T\approx 7.5~\text{h}$, $T_2 = 0.8\,T\approx 10~\text{h}$.}
\label{delay2}
\end{figure}

Figure~\ref{delay1} compares the dynamics of the supply process of an affected city $i$ when (1) the complete information about the system state as a whole is available from the beginning  and (2) the information about the state of \textit{other} cites is delayed. When all the information is available (the ideal case) the rate of the supply process increases monotonously (with saturation) and two stages can be singled out. At the first stage the number of cities from which the resource flow has gotten the city $i$ is growing and, as a result, the supply rate is increasing. At the second stage the pattern of the cities involved into the supply process becomes stable and the supply rate practically does not change in time (curve 1). When some other cities provide an insufficient information about their state (the scenario of information delay) the priority $S$ of the cities with sufficient information becomes higher. The latter leads to the increased rate of their supply as it is represented by curve 2 in Fig.~\ref{delay1}. When, however, the new information comes, the priority could be changed and the rate of supply of the city $i$ would decrease. This change of priorities gives rise to the wave-like growth. It is worthwhile to attract attention to the found fact that if the sufficient information is provided not too late after the beginning of the resource redistribution, the supply process will end practically at the same time as in the ideal case. Thereby if the resource delivery was initiated only after gaining the complete information, then the supply process would be finished with the delay $T_2$. In contrast, in the analyzed situation its duration is not affected at all due to the cooperative effects in the resource redistribution. 

The obtained results allows us to assert that the effect of the information delay on the duration of the supply process is weak due to the cooperative behavior of the resource redistribution system. 
%
To justify this statement let us analyze in more detail how the information delay affects the supply process.

Figure~\ref{delay2} shows series of plots illustrating the supply dynamics for different time moments of the information update. For all the three plots the time difference $T_2-T_1 = 0.2T$ is the same but the first time moment $T_1$ takes three different values, $T_1 = 0.2T,\ 0.4T,\ 0.6 T$.  In these plots curve 1, as previously,  
illustrates the dynamics of the supply process for an affected city $i$ if all the information is available from the beginning of the process. 
Curve 2 shows the supply dynamics for the same city $i$ when it provided the whole information about its state from the beginning of the process but there are other affected cities for which the information about their states was initially insufficient. Curve 3 depicts the supply dynamics for an affected city $j$ that provides initially an insufficient information updated at the time moments $T_1$ and $T_2$. Because at the initial step its priority was lower than the priority of the city $i$, i.e., $|S_j| < |S_i|$, the resource flow was directed to the city $i$ and only after the priorities of the cities $i$ and $j$ have become equal the resource flow was directed to the city $j$ also. After the information update the priority of the city $j$ grows in a stepwise way and, as a result, all the resource quanta are temporally sent to this city. It explains the ``screening effect'' of the city $j$ at the beginning of the supply process.  
As seen in Fig.~\ref{delay2} even in the wost case (plot III) for which the last moment of the information update $T_2 = 0.8T$ is approximately equal to the duration $T$ of the supply process in the ideal case, the constructed algorithm is able to govern the resource redistribution in such a way that the duration of the analyzed supply process actually is not affected by the information delay. 

Summarizing the obtained results, we can state that the cooperative mechanism of resource redistribution endows the system with a high adaptability and the information delay practically does not affect the duration of the recovery even in cases of comparably late collection and provision of information about the system state.

\subsection{Scenario of information incompleteness}

\begin{figure}
\begin{center}
\includegraphics[width=1.0\columnwidth]{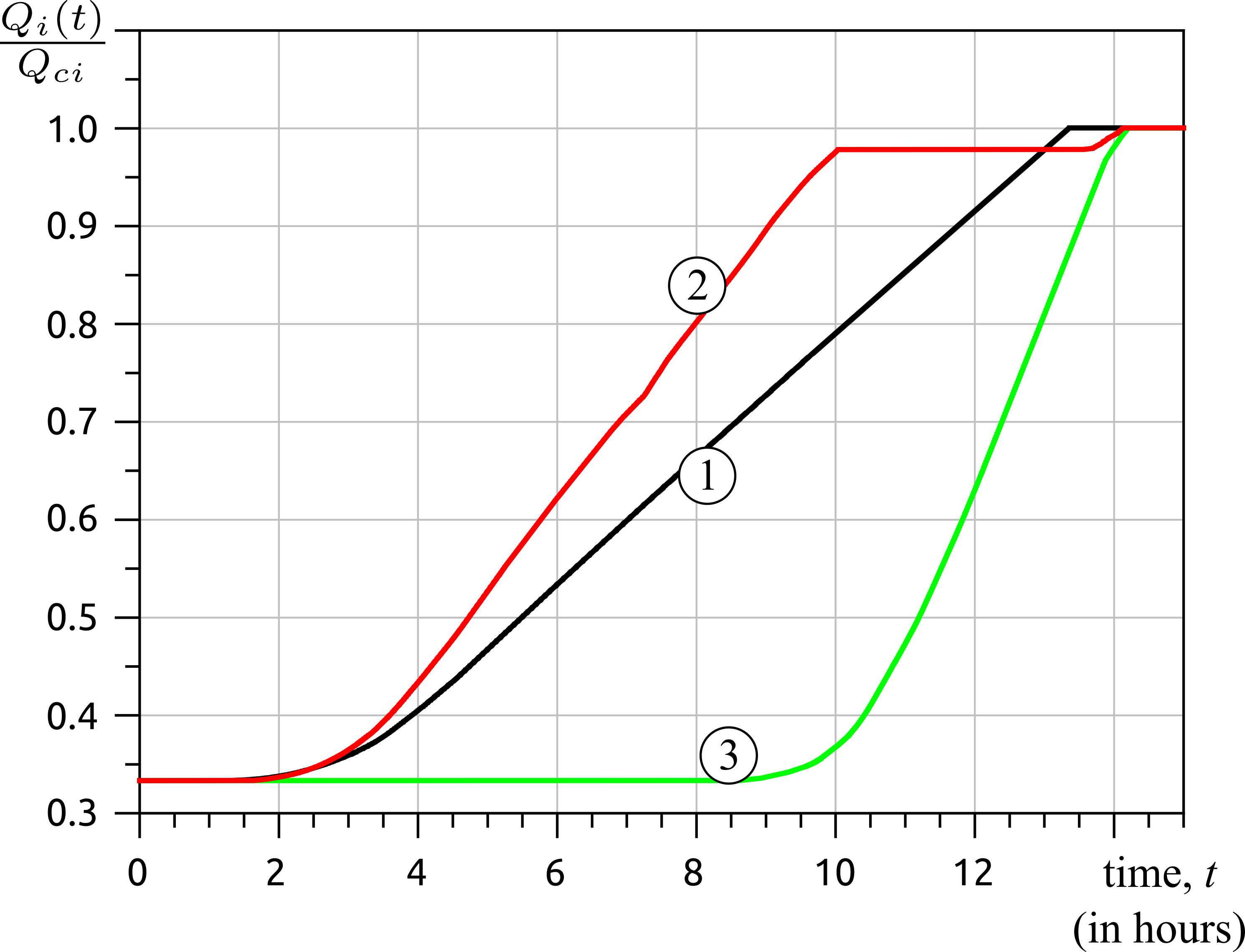}
\end{center}
\caption{Supply dynamics of affected cities in the scenario of the information incompleteness. Curve 1 matches the ideal case, curve 2 corresponds to a city $i$ that provided the sufficient information about its state from the beginning of the resource redistribution, curve 3 represents these data for a city $j$ that provides the information about its state only after the moment $T_2$ and regarded as a ``lost'' city at the beginning of the process.}
\label{incomplete1}
\end{figure}

Now let us consider the results of simulation for the system with incomplete information about the city states. As noted in Section~\ref{moddetail} we studied the case when three ones of the nine affected cities do not provide any information about them at the beginning; the same is the case with respect to four non-affected cities neighboring to the damaged region. So the process starts without them. At the time moment $T_1$ the information about these non-affected ``lost'' cities become available and they are involved into the resource redistribution. At the time moment $T_2$ the information about the affected ``lost'' cities becomes also available and the required resources are directed to the affected cities too. 

The results of simulation are presented in Fig.~\ref{incomplete1}. As previously, curve 1 shows the supply dynamics of an affected city $i$ in the ideal case. Curve 2 illustrates the supply dynamics for the same affected city $i$ the information about which was available from the beginning. Curve 3 exhibits similar data for an affected city $j$ the information about which becomes available only after the time $T_2$.
Before the time moment $T_2$ the resources were distributed over the ``visible'' cities (curve 2), so their states have become better than that of the ``lost'' cities. Therefore after the moment $T_2$ they receive the highest  priority (Table~\ref{T1}) and the resource flow is mainly directed to them. In Fig.~\ref{incomplete1} it is manifested in a very fast growth of curve~3 and a temporal stagnation in the supply of the ``visible'' cities (curve~2). On the contrary,  before the moment $T_2$ the supply of the ``visible'' cities with resources is also rather fast because no resource quanta were sent to the ``lost'' cites. In other words, broadly speaking, the ``visible'' cities are  supplied during the first part of the process, whereas the ``lost'' cities are mainly supplied during the second part. As it must be, the supply process ends practically simultaneously for all the affected cities. With respect to the ``visible'' cities the latter fact is demonstrated by a short fragment of the resource amount growth before the value $Q_i$ gets its required level $Q_{ci}^\text{real}$ (curve 2). This type of dynamics can be characterized as a certain ``screening'' effect. It should be noted that a similar ''screening'' effect but of another nature has been found in the resource redistribution during the short-term recovery in systems without the lack of information \cite{lubashevskiy2013algorithm}.

In contrast to the previous scenario, in the given case there are conditions when the resource redistribution splits into two independent processes as illustrated in Fig.~\ref{incomplete1}. Because at the initial steps three of nine cities did not provide any information about their states, no resource quanta were sent to them up to the second information update at $T_2$. Therefore the resource flow to the other six cities was substantially increased for a relatively long time interval in comparison with the scenario of the information delay. Thereby in the case of $T_1=0.4 \,T$  and $T_2= 0.6 \,T$ the system recovery is rather closed to be split into two independent supply processes (Fig.~\ref{incomplete1}). 

We can conclude this Section as follows. The results of simulation of the resource redistribution system applied to the short-term recovery after a large scale disaster shows the fact that the proposed mechanism enables a rather adaptive response to the effects of lack of information. In particular, the duration of the short-term recovery turns out to be practically insensitive to various  types of information delay if its value does not exceed the duration of the recovery process in the ideal case.  

\section{Conclusion}

This work has analyzed the effects of the lack of information on the resource redistribution process aimed at mitigating aftermath of a large scale disaster in the frameworks of short-term recovery. A socio-technological system at hand was regarded as a collection of cities connected with one another by a transport network. The effect of disaster on a given city is simulated as a lack of vital resources in it. The recovering process is based on the cooperative supply to the damaged region by the non-affected cities. The developed mechanism governing this resource redistribution generates a certain semi-optimal plan. It is based on the algorithm developed previously \cite{lubashevskiy2013algorithm} and generalized it to take into account possible effects of the lack of information or its delay.

Strictly speaking, the optimal plan of the resource redistribution implies the minimal time of the resource supply to the affected cities. One of the criteria of the plan optimality is the requirement that the resource supply be finished for all the affected cities practically at the same moment. Due to condition~\eqref{eq:order} determining the priority of the city with the worst situation the latter requirement holds. The choice of the suppling city specified by expressions~\eqref{eq:dsearch} and \eqref{eq:decrD} actually minimizes the duration of the resource redistribution. Naturally this argumentation is not a rigorous proof of optimality. However, it meets the necessary criteria, enabling us to call it semi-optimal.

Using numerical simulation we have studied the influence of the lack of information on the resource redistribution planning. Two cases of the lack of information were considered. The first one is the delay of the information about the state of the affected cities. The second one is the information incompleteness with respect to affected and not-affected cities.

As a main result, we have demonstrated that the lack of information affects weakly the duration of the recovering process as a whole. Besides, it is shown that the delay of information as well as its incompleteness manifest themselves in local screening effects.   











\bibliography{mylibrari}

\end{document}